# The Galactic Cosmic Ray Electron Spectrum from 3-70 MeV

# Measured by Voyager 1 Beyond the Heliopause – What This

# Tells Us About the Propagation of Electrons and Nuclei In and Out

# of the Galaxy at Low Energies

W.R. Webber and T.L. Villa

New Mexico State University, Astronomy Department, Las Cruces, NM 88003, USA

**ABSTRACT**

The cosmic ray electrons measured by Voyager 1 between 3-70 MeV beyond the heliopause have intensities several hundred times those measured at the Earth by PAMELA at nearly the same energies. This paper compares this new V1 data with data from the earth-orbiting PAMELA experiment up to energies >10 GeV where solar modulation effects are negligible. In this energy regime we assume the main parameters governing electron propagation are diffusion and energy loss and we use a Monte Carlo program to describe this propagation in the galaxy. To reproduce the new Voyager electron spectrum, which is ~$E^{-1.3}$, together with that measured by PAMELA which is ~$E^{-3.20}$ above 10 GeV, we require a diffusion coefficient which is ~$P^{0.45}$ at energies above ~0.5 GeV changing to a $P^{-1.00}$ dependence at lower rigidities. The entire electron spectrum observed at both V1 and PAMELA from 3 MeV to 30 GeV can then be described by a simple source spectrum, dj/dP ~$P^{-2.25}$, with a spectral exponent that is independent of rigidity. The change in exponent of the measured electron spectrum from ~-1.3 at low energies to ~3.2 at the highest energies can be explained by galactic propagation effects related to the changing dependence of the diffusion coefficient below 0.5 GeV, and the increasing importance above 0.5 GV of energy loss from synchrotron and inverse Compton radiation, which are both ~$E^2$, and which are responsible for most of the changing spectral exponent above ~1.0 GV.

As a result of the $P^{-1.00}$ dependence of the diffusion coefficient below 0.5 GV that is required to fit the V1 electron spectrum, there is a rapid flow of these low energy electrons out of the galaxy. These electrons in local IG space are unobservable to us at any wave length and therefore form a "dark energy" component which is ~100 times the electrons rest energy.

## Introduction

The Voyager 1 (V1) high precision measurement of the spectra of GCR nuclei from H to Fe as well as electrons beyond the Heliopause (HP) by the CRS instrument (Stone, et al., 2013), is one of the triumphs of modern technology and speaks well for the prescience of the instrument designers. The instrument was designed to measure what were estimated in the 1970's to be the local interstellar spectrum (LIS) of cosmic ray nuclei and electrons at a time when the HP location was estimated to be at only ~5-10 AU and with a very limited idea of what the LIS intensities really were (Stone, et al.., 1977). The features and intensities of these components that are now observed at V1, which match so well to those calculated prior to the HP crossing (e.g., Webber and Higbie, 2008, 2009) form some of the strongest evidence that V1 has indeed been in local galactic space since August 25$^{th}$, 2012.

The electron intensities that were measured by V1 after the HP crossing at 121.7 AU (Stone, et al., 2013) are several hundred times the intensities that have been recently measured by PAMELA down to below 100 MeV near the Earth in 2009 at a time of a grand minimum in solar modulation (Bozio, et al., 2014). It is therefore not simple to put this new low energy measurement in perspective in terms of heliospheric modulation, when, at higher energies, the PAMELA experiment has also measured intensities of electrons at ~1 GeV in 2009, that were only a factor 3-4 lower than the estimated LIS electron intensities at the same energies (Potgieter, 2013). At 10 GeV and above these same PAMELA intensities are within 10-20% of the estimated LIS intensities and are, in fact, used to define the LIS spectrum at these higher energies.

It is our objective in this paper to connect the low energy Voyager measurements to the PAMELA measurements of electrons at higher energies near the Earth and create an accurate LIS electron spectrum over the entire energy range from a few MeV to above 10 GeV where solar modulation effects are neglible. We will use a simple Monte Carlo diffusion model for electrons in the galaxy (Webber and Rockstroh, 1997). This model starts with a source rigidity spectrum of the form $dj/dP \sim P^{-x}$, where x is a constant exponent throughout the rigidity (or energy) range involved. All of the observed changes in the electron spectrum after it is initially accelerated (its "source" spectrum) are produced by "propagation" effects related to the diffusion and energy losses of these electrons as they propagate to us.

**The Data**

The Voyager 1 electron data (Stone, et al., 2013; Cummings, et al., 2016) are shown in a j x $E^2$ format in Figure 1 along with the recent PAMELA data at Earth (Bozio, et al., 2014) which now extends from less than 100 MeV all the way up to 100 GeV and above. This PAMELA data illustrates the strong effects of solar modulation up to a few GeV. This figure shows the intensities times $E^2$, thus showing the differences in intensities at lower and higher energies on a more comparable scale.

At an energy ~80 MeV the PAMELA measurements and those from Voyager 1 beyond the HP differ by a factor of 500-1000. Unfortunately, the measurements do not quite overlap in energy so this is a projection at 80 MeV, mid-way between the upper V1 energy and the lowest PAMELA energy. At this energy the effects of Jovian electrons are believed to be negligible, so this difference in intensities represents the effects of solar modulation between Voyager and PAMELA.

At low energies the new V1 interstellar data is consistent with an electron spectrum ~$E^{-1.3}$ (Cummings, et al., 2016). Above a few GeV the PAMELA data, Adriani, et al., 2013, and other high energy data from FERMI, Ackermann, et al., 2012, and AMS-2 (Aguilar, et al., 2014) are consistent with a spectrum ~$E^{-3.1-3.2}$ above 10 GeV. So that, to fit the observed LIS electron spectrum at both the lowest and highest energies, requires propagated LIS spectra that differ by 1.7-1.8 in the exponent between low and high energies.

**Propagation of Electrons in the Galaxy**

The Monte Carlo diffusion model (MCDM) used here is described more fully in Webber and Rockstroh, 1997, and Webber and Higbie, 2008. For the electron specific calculations we assume that the magnetic field B=$B_0$ exp$^{-Z/Z_B}$ where $Z_B$=1.5 Kpc. The value of $B_0$= 6μG, so that the energy loss by synchrotron radiation, dE/dt = -3.7x$10^{-16}$ $E^2$ GeV/sec at Z=0. Similarly we have for the energy loss from the inverse Compton effect, dE/dt = -1.3x$10^{-16}$ $E^2$ GeV/sec with $Z_{IC}$ = 3 Kpc.

The energy loss by bremsstrahlung and ionization is determined from the matter distribution. Here we take the interstellar density $n_0$=0.6 cm$^{-3}$ at $Z_0$. Thus, dE/dt for Br = -1.0 x $10^{-15}$ E GeV/nuc at $Z_0$. The Z dependence of the total matter density is $\sim\exp^{-Z/Zm}$ where $Z_m=0.2$ Kpc. This gives a total matter line integral $=\int ndz = 4.5\times10^{20}$ cm$^{-2}$.

The diffusion coefficient is given by K (P) = $2\times10^{28}$ cm$^2$/sec$^{-1}$ at 1 GV at Z=0. For the "thick disk" diffusion thickness we use $Z_L \pm 1.0$ Kpc. This leads to an electron lifetime $\tau=L^2/2K$ at 1 GV, of 1.2 x $10^7$ yr. This is comparable to the measured GCR lifetime for nuclei obtained using the $^{10}$Be isotope which is determined to be =1.5 x $10^7$ year at $\sim$1.5 GV (Yanasuk, et al., 2001). These parameters are very similar to those used in the earlier Webber and Higbie, 2008, calculations for electrons using the same diffusion model.

The relative importance of the various loss terms on the electron spectrum is shown in Figure 2. It is significant that the loss terms which are $\sim E^2$ become the dominant loss terms above $\sim$0.5 GV and this $E^2$ dependent loss means that the value of negative exponent of the source spectrum will be increased by nearly 1.0 power in the exponent between 0.5 and 10 GV as a result of interstellar propagation due to this effect alone (see also Figure 6 showing the calculated spectral exponent vs. energy for electrons).

The electrons are all injected at Z=0 in the model and the resulting final distribution is binned in both the Z dimension and in energy. The total integration time of the calculation is $\sim$2 times the characteristic diffusion life-time of $L^2/2K$ to the boundary at all energies thus resulting in an equilibrium distribution of electrons at each energy and location.

### Discussion of the Propagation Results

We first calculate the LIS spectra starting with source spectra with exponents S = -2.2-, -2.3 and -2.4. These "LIS" spectra were calculated using a diffusion coefficient which is $\sim P^{0.45}$ at rigidities above a value of $P_0 = 1.0$ GV and equal to a constant = 2.0 x $10^{28}$ cm$^2\cdot$s$^{-1}$ below 1 GV. The value of this diffusion coefficient as a function of P is shown as a black curve in Figure 3. The resulting LIS spectra for S= -2.2, -2.3 and -2.4 are shown in Figure 4. The spectra are all normalized to an intensity $j_0$=1.5 x $10^2$ e/m$^2\cdot$sr$\cdot$GV at 1 GV (see Webber and Higbie, 2008). The calculated LIS spectra for S=-2.30 $\pm$ 0.05 provide a reasonable fit to the PAMELA spectra between 3-30 GeV when the effects of solar modulation at lower energies are included (see also Potgieter, et al., 2013), with small adjustments of the value of $j_0$ to fit the different source spectra to the PAMELA data.

The assumption of K (P) equals a constant below ~1 GV gives intensities that are too large, however, and do not have the right slope at the lower energies where the V1 electron spectrum is measured.

In order to fit both the V1 data and PAMELA data simultaneously and maintain a source spectral exponent which is constant with rigidity we need to assume a break in the exponent of the rigidity dependence of K (P). Such a break at ~1.0 GV or below has been predicted and discussed earlier by Ptuskin, et al., 2006.

The lower energy part of the LIS electron spectra are calculated using galactic diffusion coefficients of K which are ~$P^{-1.0}$ below $P_0$. For values of $P_0$ = 1.00, 0.562, 0.427, 0.316 and 0.176 GV and for a source spectrum ~$E^{-2.25}$ these spectra are shown in Figure 5. The $P^{-1.0}$ dependence gives spectra at ~50 MeV and below that are ~$E^{-1.30}$ for all of the above examples of $P_0$, but the intensities match those measured at V1 only for $P_0$ between 0.427 and 0.562 GV, thus fixing this parameter at about 0.5 GV. Figure 5A shows an expanded version of the data and the calculations below ~100 MeV.

The observed slope of -1.3 for the electron spectrum thus fixes the slope of K (P) to be ~$P^{-1.0}$. The exponent of the source spectrum, therefore, remains almost unchanged at -2.25 from ~3 MeV up to ~10 GeV. Calculations with K (P) ~$P^{-0.5}$ below $P_0$, also with S= -2.25 for the source spectrum, give electron spectra with a slope ~-1.5 below 50 MV. This slightly larger exponent is the same exponent that was originally reported for the LIS electron spectrum (Stone, et al., 2013).

So, as a result of this break in the rigidity dependence and also the fact that the dominant energy loss terms for electrons at high energies become unimportant below about 0.5 GeV, the observed LIS electron spectral exponent changes from ~-3.1 to-3.2 at high rigidities to a slope ~1.3 at the low rigidities where the Voyager 1 spectrum is measured. This calculated changing slope of the LIS electron spectrum as a function of energy is shown in Figure 6 where S = -2.25 and $P_0$ = 0.316, 0.562 and 1.00 GV.

Thus the measured electron intensities and the electron spectral index measured at V1 provide a very sensitive test for the value of, and rigidity dependence of, the interstellar diffusion coefficient for all cosmic rays at these lower rigidities.

Note that the spread in the calculated electron intensities at ~10 MeV for a spectral break at $P_0 = 0.427$ GV, and a break at $P_0 = 0.562$ GV is a factor of about 30% as shown in Figure 5A. This difference is larger than the experimental errors in the measured intensities which are estimated to be $\pm 15\%$ (see Cummings, et al., 2016, and the data shown in Figure 5A).

The electron data can therefore be well fit at both low and high rigidities with a single source spectrum with index = ~2.25, and a break in the rigidity dependence of K at about 0.5 GV and using a rigidity dependence of $K \sim P^{-1.0}$ at lower rigidities, using only three simply connected parameters. The normalization at 1 GV is 1.47 x $10^2 e/m^2 \cdot sr \cdot s \cdot GV$. This is an intensity that is ~3.5 times that measured at 1.0 GV at PAMELA in 2006-2007. This intensity difference corresponds to a modulation parameter = 400 MV, the same as that determined for the protons at the same rigidity, also in 2006-2007.

These intensities below a few GeV from these calculations may be used to define a LIS spectrum which then may be compared with the PAMELA measured intensities to define the effects of solar modulation on this spectrum. It can be seen from Figure 5 that the calculated and measured electron spectra converge above a few GeV indicating that solar modulation effects become smaller with increasing energy.

This convergence into a single unmodulated spectrum of index = -3.1 to -3.2 at ~10 GV and above then may in turn be used to define the source spectrum to be also between ~-2.25 and -2.35 at higher energies up to 1 TeV.

## The Significance of the Negative Value of the Exponent in the K (P) Dependence of the Diffusion Coefficient at Low Rigidities

The implications of the dependence of the diffusion coefficient which could be K (P) $\sim P^{-0.5}$ or $P^{-1.0}$ at lower energies are significant. This means that most of the electrons below a few 100 MeV rapidly escape from the galaxy, never to be detected again. They carry with them a typical kinetic energy ~100 x their rest energy of 0.5 MeV. In Figure 7 we show the trapping efficiency of the galactic disk as a function of energy (this is the fraction of the $10^4$ electrons injected at each energy that remain in the galaxy). This illustrates the rapidity at which these electrons below a few 100 MV vanish into a region beyond 1 Kpc where we cannot detect them. Already at 30 MeV for $P_0$=0.562 GV, greater than 90% of the injected electrons have escaped the galactic disk.

This almost undetectable kinetic energy could be considered a source of "dark" energy, which is travelling at the speed of light (almost) and slowly filling up the space between galaxies and most likely many other earlier entities in the evolving Universe as these electrons (and nuclei) are created under much more active conditions.

## Summary and Conclusions

The Voyager 1 measurements of the GCR electron spectrum from ~3-70 MeV beyond the heliopause have permitted for the 1st time a realistic determination of the propagation and source spectrum of galactic electrons below ~1 GeV. The intensities and spectrum that are measured by Voyager and also at higher energies by PAMELA may be fit by an electron source spectrum of exponent between -2.2 and -2.3, which is independent of energy. The electron spectrum observed by V1, which has an index of about -1.3, slowly changes into an exponent of index ~-3.1 to -3.2 at high energies. This changing exponent can be explained by the fact that the synchrotron and inverse Compton processes, which depend on $E^2$ and which determine the spectral changes above ~1.0 GeV, become less important at energies below ~1 GeV; along with the effects of the changing dependence of the diffusion coefficient with rigidity below 0.5 GV. This increasing value of the diffusion coefficient below $P_0$ increases the probability of escape of the lower energy particles from the galaxy. This rapid escape has reduced the LIS spectral index that is actually observed by V1, which is ~$E^{-1.3}$, by about 1.0 power below the source index of -2.2 -2.3 at energies below ~100 MeV.

A diffusion coefficient at low energies which becomes ~$P^{-1.0}$ at ~ 0.5 GeV, along with an electron source spectrum with a constant spectral index = -2.25, and a normalization = 1.47 x $10^2$e/m$^2$·sr·s·GV at 1 GV will therefore fit the electron data measured by Voyager to within $\pm$ 10% at 10 MeV and also the PAMELA data at 10 GeV.

Ptuskin, et al., 2006, have predicted a spectral break in the galactic diffusion coefficient at ~1.0 GV or below on dynamical grounds.  According to these authors, the near equality of the energy density of the galactic B field and of the GCR which are confined by this field, may lead to instabilities in the field thus modifying the diffusion coefficient.  In their paper a very steep rigidity dependence, K ~$P^{-1.0}$ similar to the value used in this paper, was in fact, considered at low rigidities.  The Voyager electron spectrum and the intensities measured at lower energies support this earlier conjecture.

The total energy density of GCR has been determined from V1 measurements of the H, He and heavier nuclei and including electrons beyond the heliopause to be between 0.8 and 1.0 ev/cm$^3$ (Cummings, et al., 2016).  This is equivalent to the energy density in a (galactic) B field of magnitude between 5.7-6.2 μG.  The actual magnitude of the B field beyond the heliopause, also measured at V1 (Burlaga and Ness, 2014) varies from ~4 to 6 μG with an average of about 4.6 μG.  This value may be modified by the localized interaction with the heliosphere.  But the near equality of the above numbers for the energy density of GCR and the B field could support the arguments presented by Ptuskin, et al., 2006, and also supports the widely used assumptions regarding the equality of magnetic energy density and cosmic ray energy density in galaxies and in more localized acceleration regions (e.g., Beck, 1991), and a typical galactic magnetic field of magnitude ~6μG.

The energetic electrons that leave our galaxy (and energetic nuclei as well) form a reservoir of essentially undetectable mass and energy in intergalactic space, perhaps fed by even more energetic situations in the earlier evolving universe.  Perhaps our universe "initially" had a "mass", but as time has evolved, more and more of this mass and energy that was initially present became invisible to us.  This entire scenario reminds one of an old Cossack song with new lyrics by Pete Segar in 1955 and made famous by the Kingston Trio and others, "Where have all the flowers gone", only with minor substitutions as follows:

Where have all the "electrons" gone, long time ago
Where have all the "electrons" gone, long time ago
Gone to "Dark Energy", every one
Gone to "Dark Energy", every one

When will they ever learn?

When will they ever learn?

**Acknowledgements:** The authors are grateful to the Voyager team that designed and built the CRS experiment with the hope that one day it would measure the galactic spectra of nuclei and electrons. This includes the present team with Ed Stone as PI, Alan Cummings, Nand Lal and Bryant Heikkila, and to others who are no longer members of the team, F.B. McDonald and R.E. Vogt.

**FIGURE CAPTIONS**

**Figure 1:** Measurements of the GCR electron spectrum between 3-70 MeV made at V1 beyond the heliopause at ~122 AU (Stone, et al., 2013; Cummings, et al., 2016), and those made by PAMELA in 2009 up to above 10 GeV (Bozio, et al., 2014), at a time of minimum solar modulation at the Earth. The black vertical line at 80 MeV represents a solar modulation factor = 700.

**Figure 2:** The various energy loss or escape processes affecting the spectrum of GCR electrons during their propagation in the galaxy. The line labelled Sync and IC represents the sum of the synchrotron and inverse Compton loss, both of which are ~$E^2$. The line labelled Br is the Bremsstrahlung loss which is ~E and ion is the ionization loss which is almost independent of energy. The parameters used to calculate these quantities are described in the text. The red line represents the diffusion escape of electrons out of the galaxy, which depends on the value of K (P) and the boundary distance assumed. This lifetime is "normalized" to the E/(dE/dt) "lifetime" of the energy loss process by using the measured lifetime of cosmic ray $^{10}$Be nuclei at 1-2 GV of 1.5 x $10^7$ yr., as obtained by Yanasuk, et al., 2001

**Figure 3:** The diffusion coefficient K (P) as a function of rigidity, P. The black line is an earlier dependence used by Webber and Higbie, 2008. The red lines represent possible modifications to this earlier dependence required to fit both the V1 and PAMELA electron data simultaneously.

**Figure 4:** The same measurements for electrons as shown in Figure 1 with the addition of LIS spectra calculated using a Monte Carlo Diffusion Model (e.g., Webber and Higbie, 2008). The source rigidity spectra for these calculations have exponents = -2.4, -2.3 and -2.2 and the diffusion coefficient is ~$P^{0.5}$ above 1 GV. Below 1 GV the diffusion coefficient is independent of rigidity as shown by the black line in Figure 3. All three electron spectra are normalized at 1 GV where the intensity is taken to be 1.5 x $10^2$ electrons/$m^2$.sr·GeV (see Webber and Higbie, 2008). These calculated intensities all exceed those measured by V1, but would agree with the intensities measured by PAMELA for S=2.30 $\pm$0.05.

**Figure 5:** Similar to Figure 4 except the calculated LIS for electrons are now modified by changing the rigidity dependence of the galactic diffusion coefficient below 1 GV as illustrated by the red lines in Figure 3. The source rigidity spectrum of electrons for the calculations in Figure 5 has an exponent = -2.25. Below $P_0$=1.0 GV, 0.562, 0.427, 0.316 and 0.176 GV the dependence of K is ~$P^{-1.00}$.

**Figure 5A:** Expanded version of Figure 5 showing energy region below 100 MeV and providing a better sense of the experimented and calculated errors.

**Figure 6:** The spectral index as a function of energy of the LIS electron spectra for source spectra with S= -2.25 and for $P_0$= 0.316, 0.562 and 1.00 GV. Because of synchrotron and inverse Compton losses above 1-2 GV and diffusion losses between $P_0$ and 1-2 GV, this LIS index slowly decreases from -3.2 at ~10 GV to about -2.3 just below 1.0 GV before plunging at $P_0$ as the diffusion index changes, so as to asymptotically approach an index -1.3 for all values of $P_0$ below about~0.1 GV.

**Figure 7:** The trapping fraction for electrons in the galaxy as a function of energy for different values of $P_0$. At ~1 GeV, ~30% of all electrons injected never leave the galaxy. At 0.1 GeV and 10 GeV the number of electrons remaining in the disk is only ~30%. At 10 MeV and below, less than 10% of the electrons injected are still in the disk.

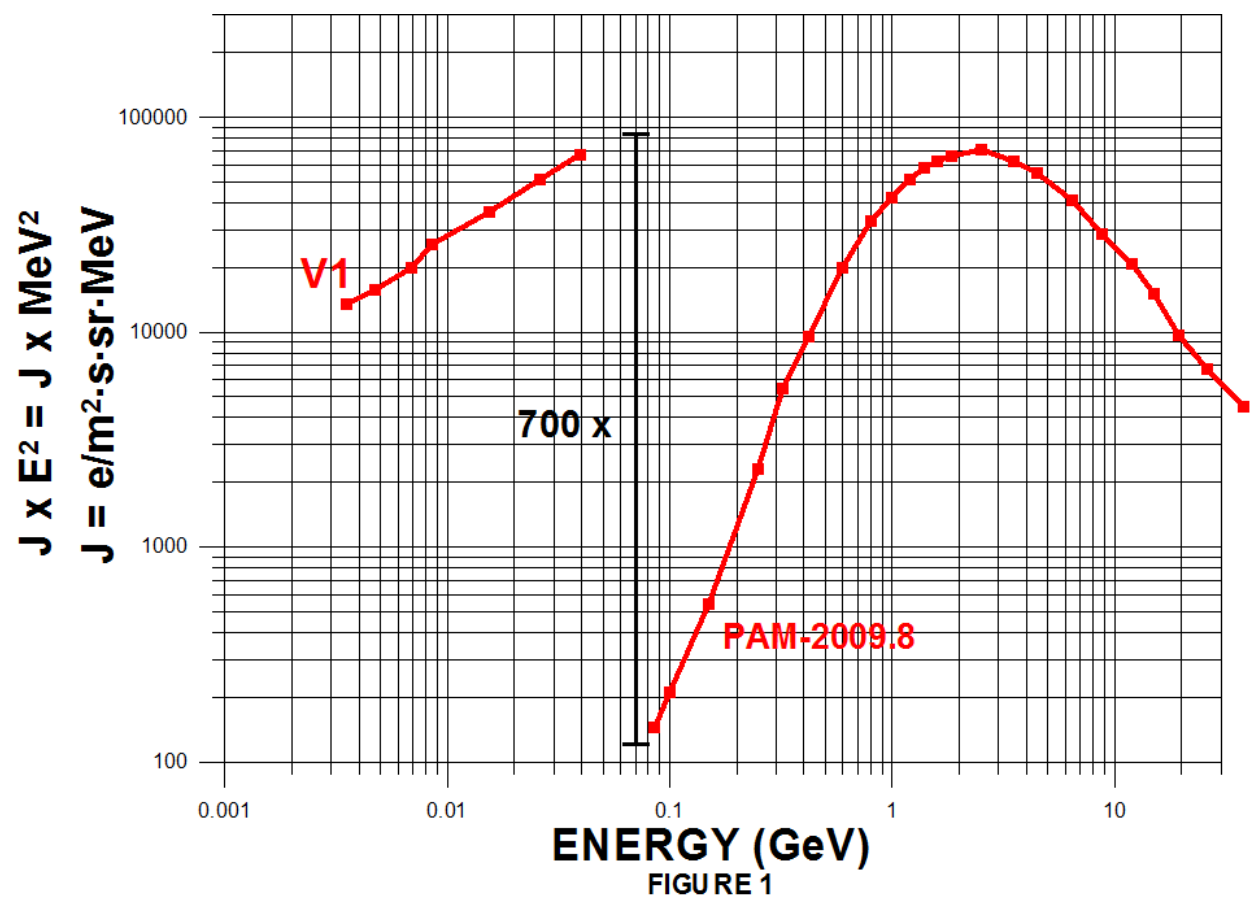

**FIGURE 1**

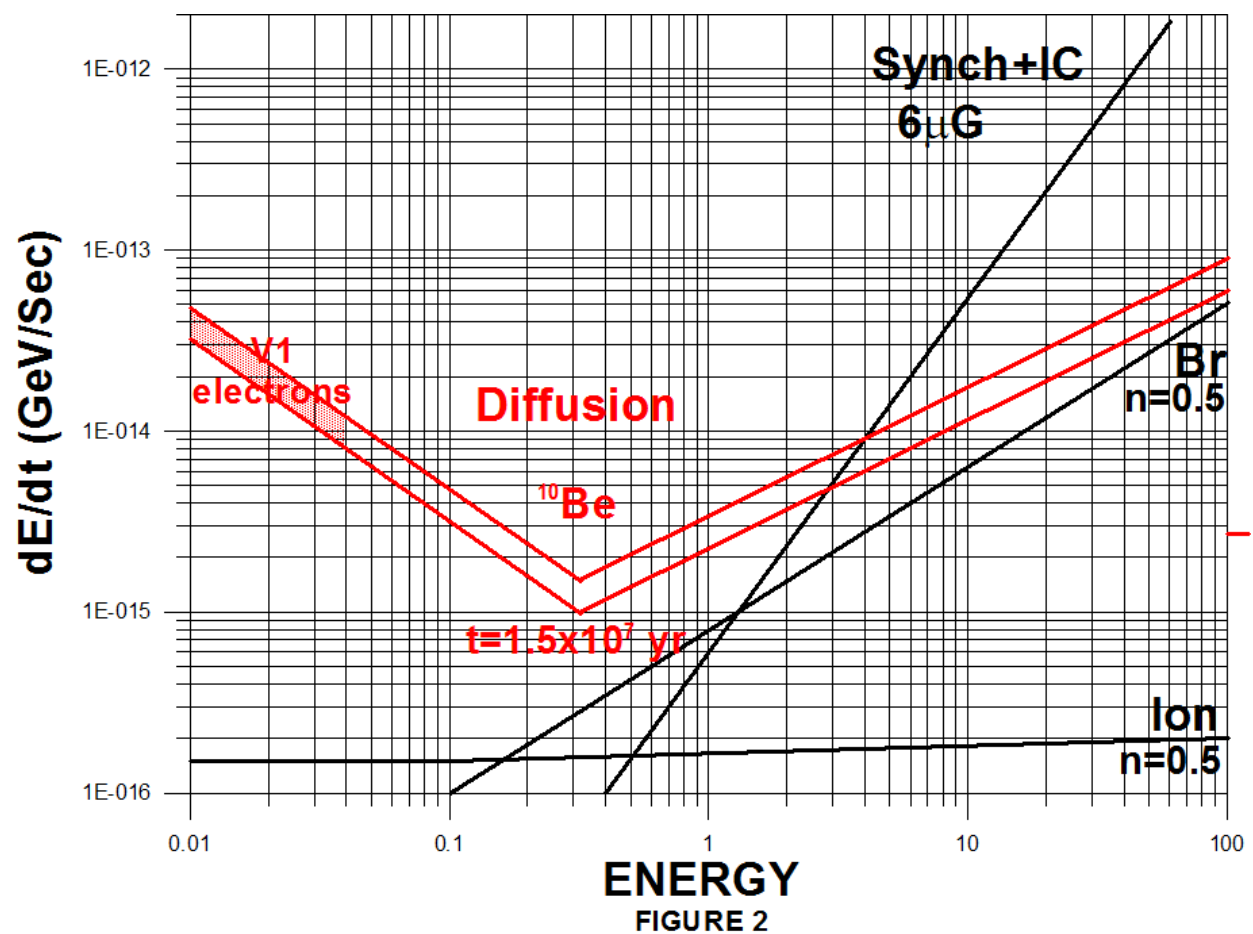

**FIGURE 2**

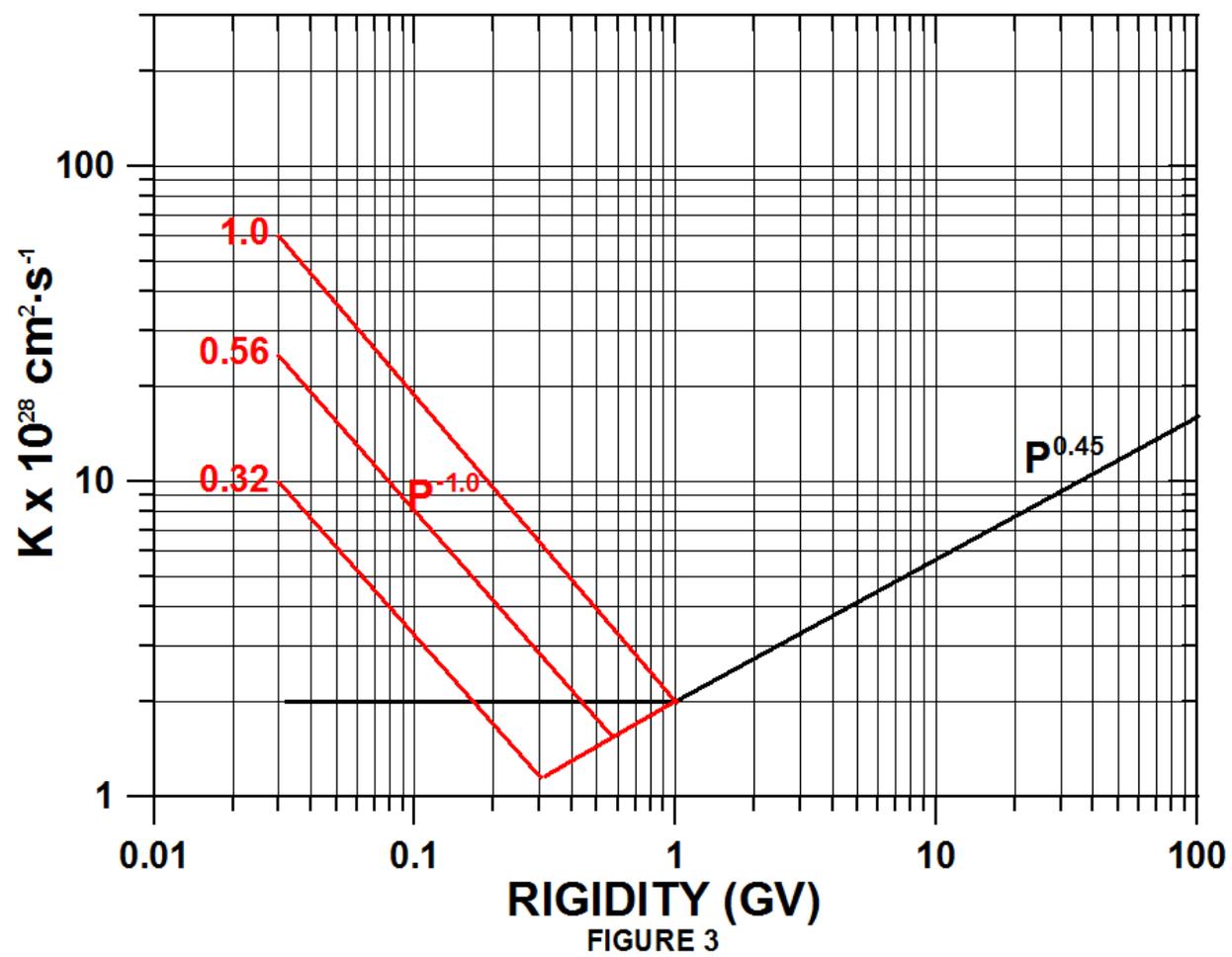

FIGURE 3

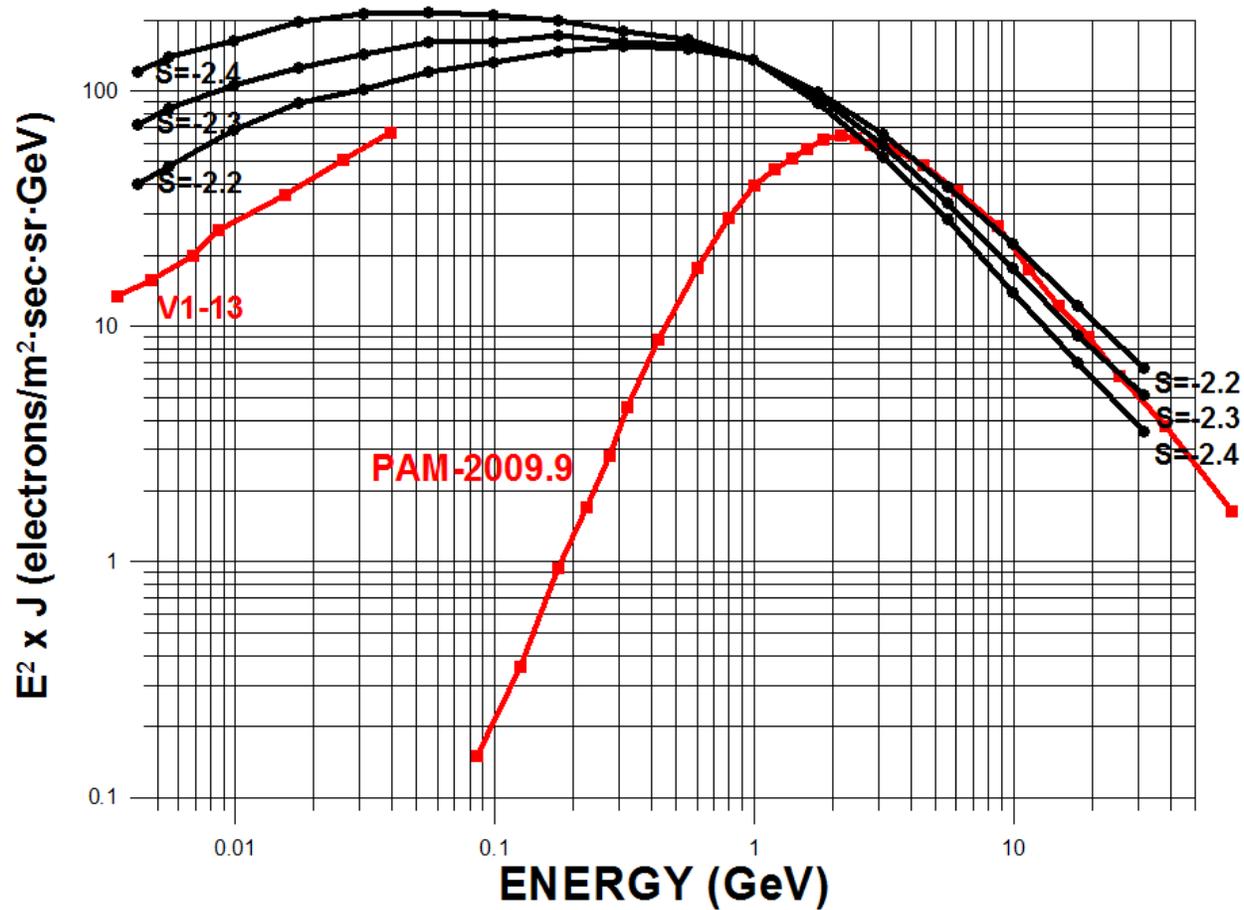

FIGURE 4

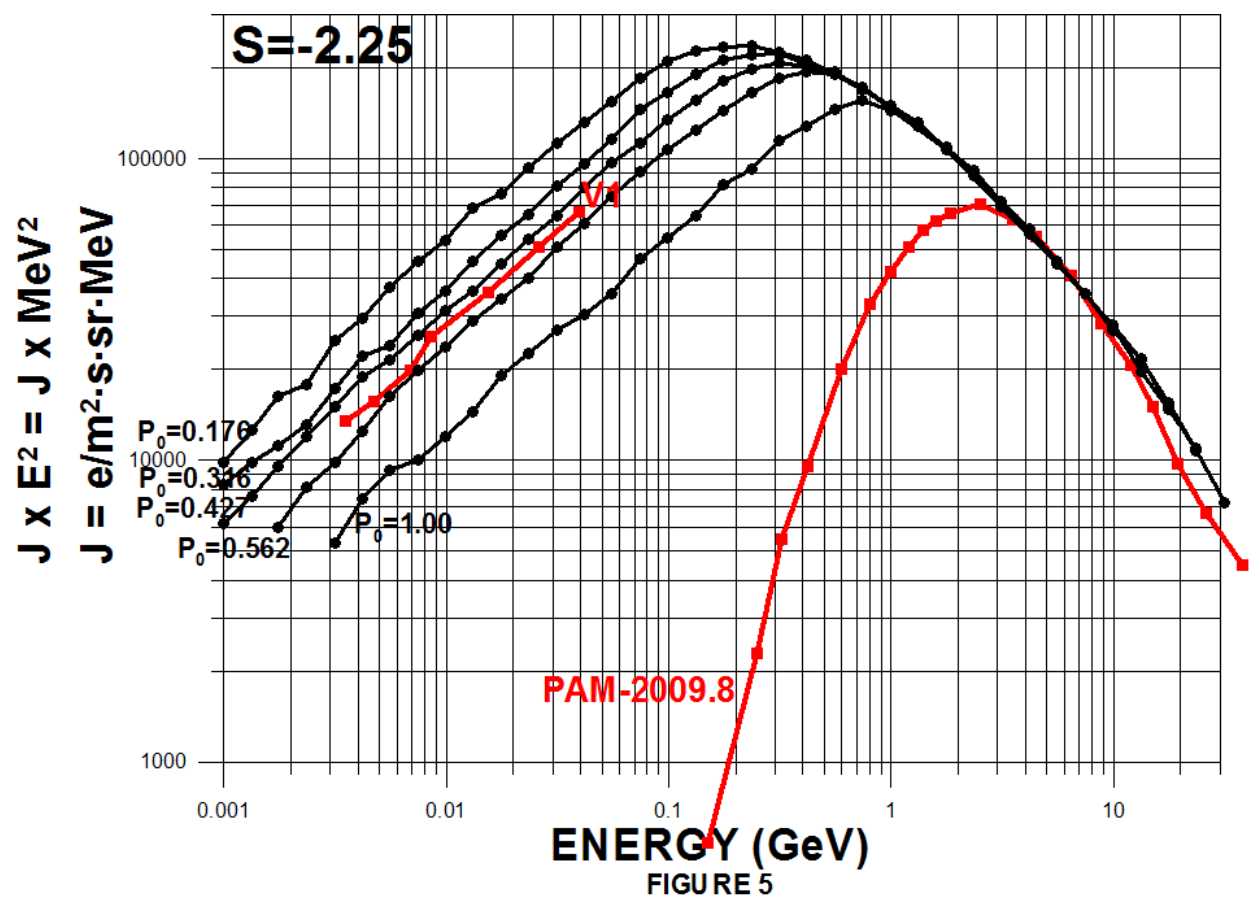

FIGURE 5

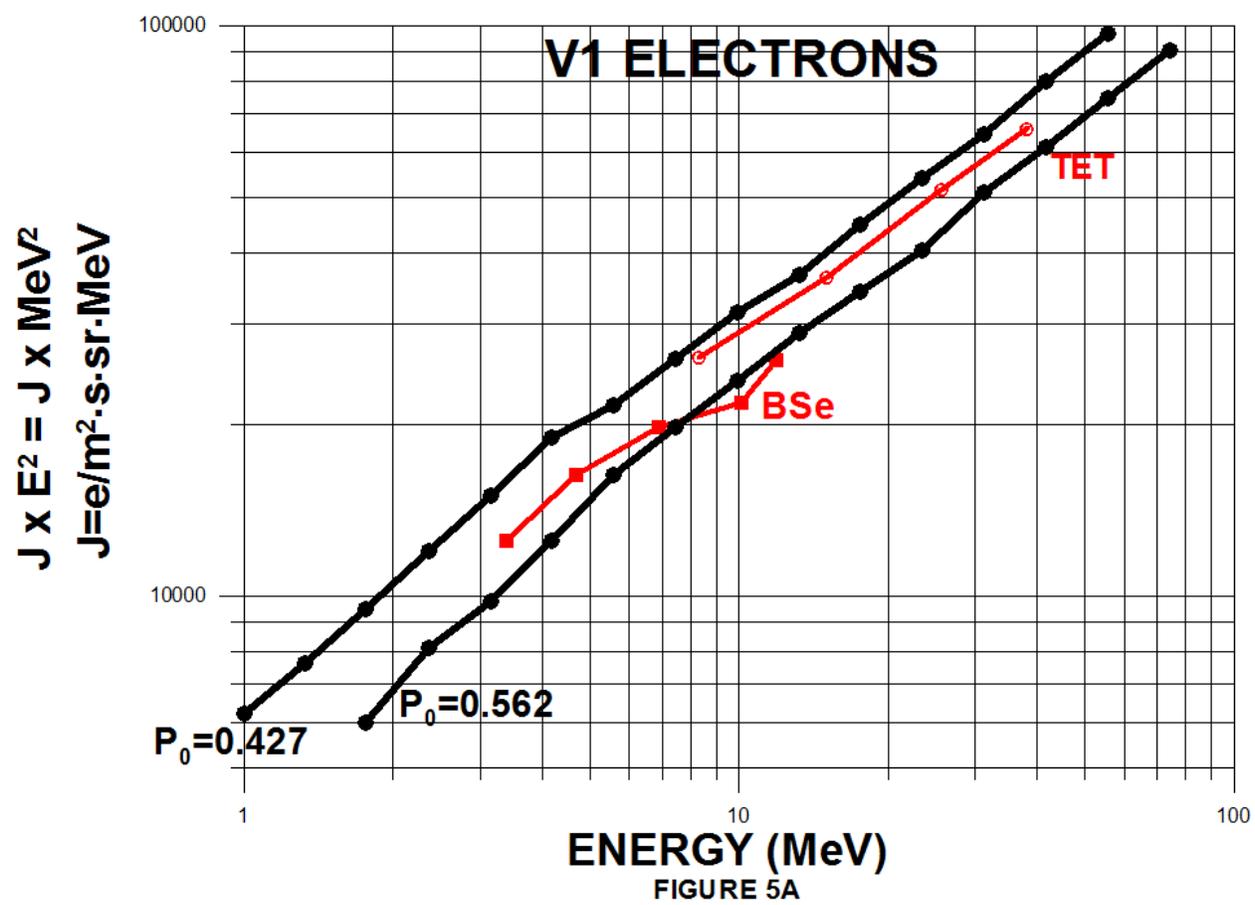

FIGURE 5A

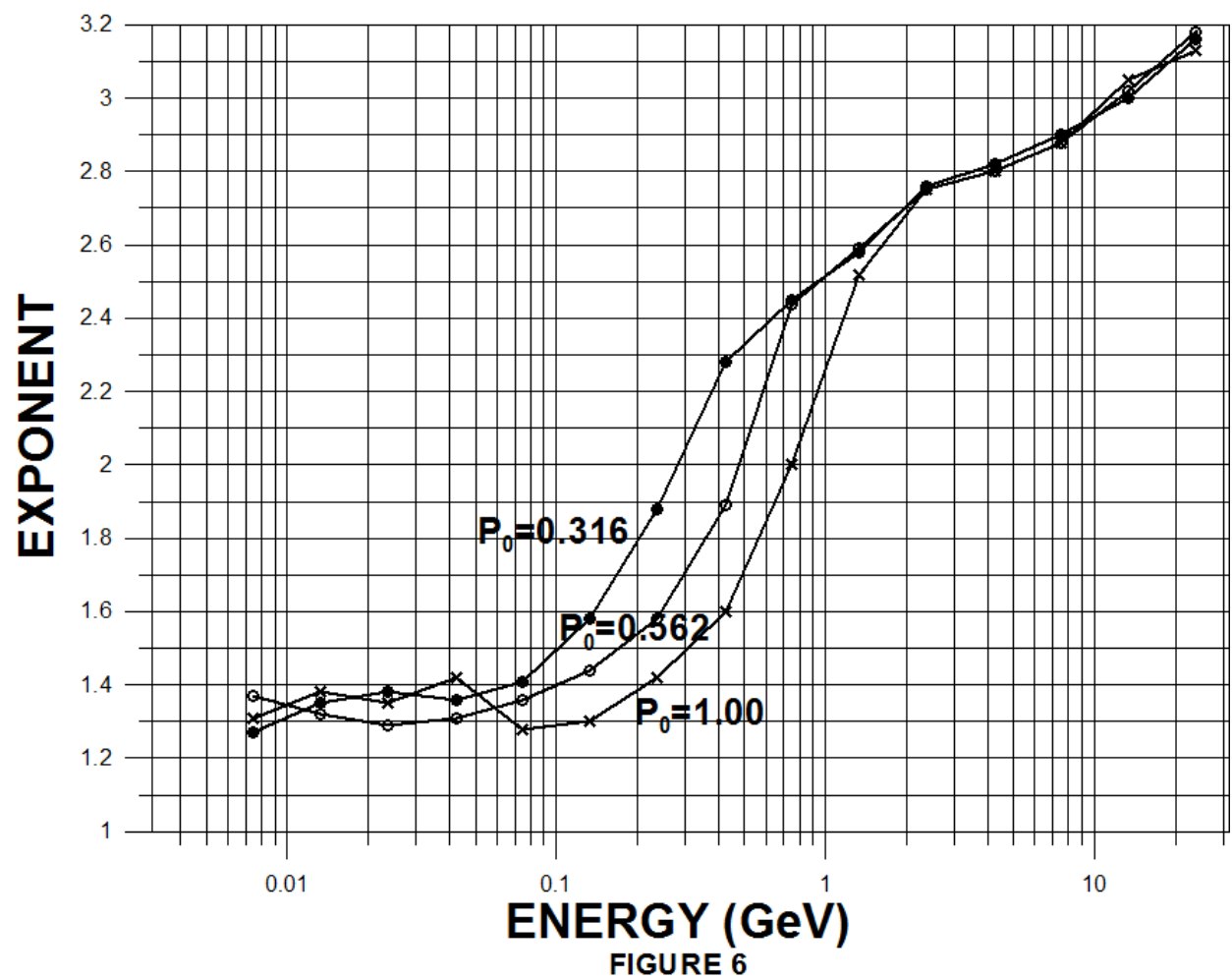

**FIGURE 6**

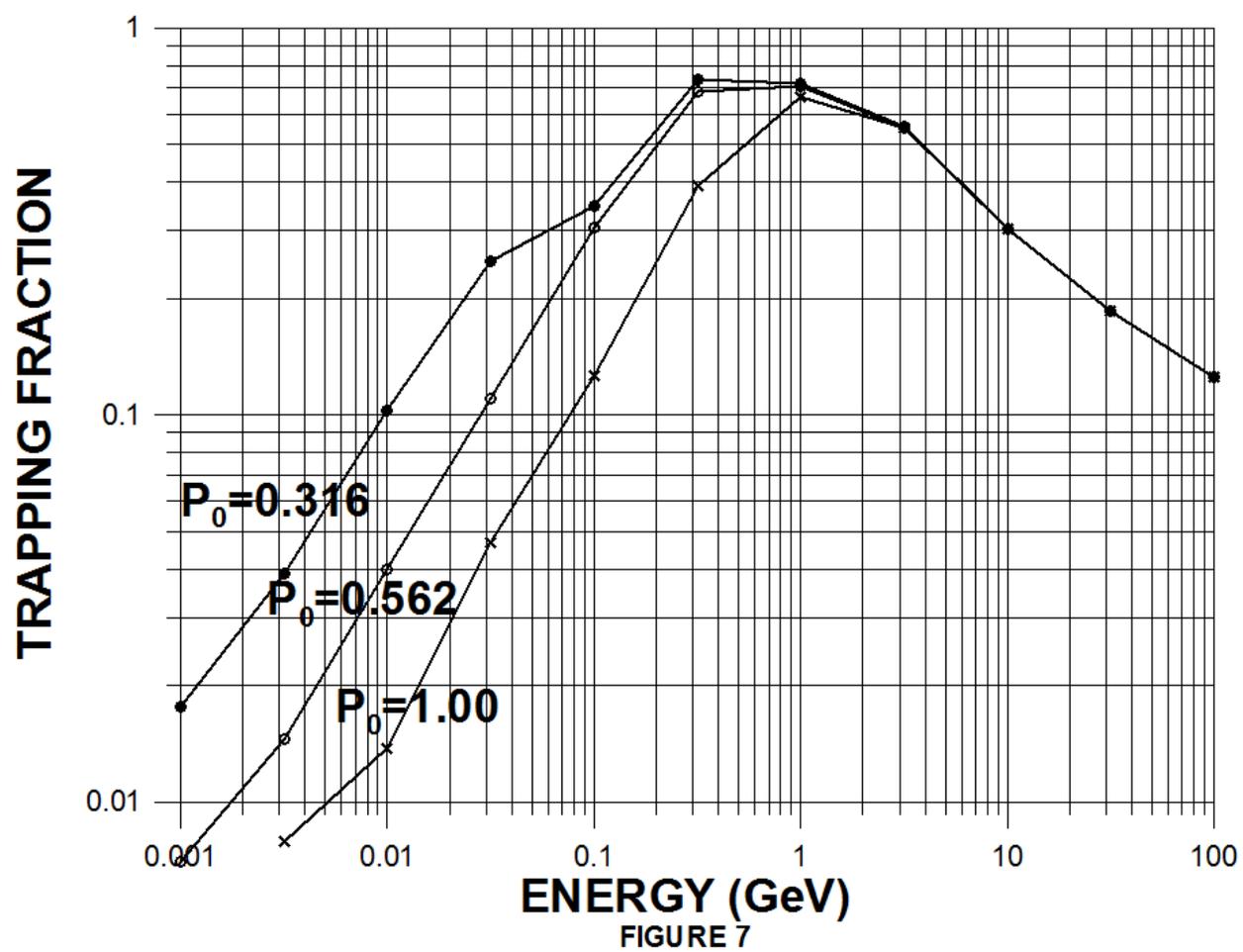

FIGURE 7